\documentclass{article}
\usepackage{spconf,amsmath,graphicx}
\usepackage{tabularray}
\usepackage{color,soul}
\usepackage{graphicx}
\usepackage{subcaption}
\usepackage{pdfpages}
\usepackage{soul} % for st{}
\usepackage{multirow}
\usepackage{subcaption}
\usepackage{hyperref}
\usepackage{amsfonts}

\title{Locality enhanced dynamic biasing and sampling strategies for contextual ASR}
\name{Md Asif Jalal$^1$, Pablo Peso Parada$^1$, George Pavlidis$^2$,Vasileios Moschopoulos$^2$,\\ Karthikeyan Saravanan$^1$, Chrysovalantis-Giorgos Kontoulis$^2$, Jisi Zhang$^1$, Anastasios Drosou $^2$,\\  Gil Ho Lee$^3$, Jungin Lee$^3$, Seokyeong Jung$^3$}
\address{
  $^1$Samsung Research UK, United Kingdom,
  $^2$Centre for Research and Technology Hellas, Greece,\\
  $^3$AI R\&D Group, Samsung Electronics, Suwon, South Korea}
\copyrightnotice{979-8-3503-0689-7/23/\$31.00~\copyright2023 IEEE}
\begin{document}
%\ninept
%
\maketitle

\begin{abstract}
% The success of Automatic Speech Recognition(ASR) depends on the underlying ASR model’s ability to recognize rare phrases relevant to the individual user and time variant. Contextual biasing~(CB) modules bias ASR model’s towards contextually-relevant phrases which may not be known until inference time. Such phrases include but are not limited to the user’s contact names, installed applications, contents of their media library, and local points of interest.  Among many approaches, previous work has performed CB by biasing the ASR model's internal representation as well as the output posteriors as language model (LM) fusion. In this paper, we further distil the contextual biasing vector from text-acoustic fusion of a given acoustic representation by using its locality information and projecting the resulting bias vector on the ASR model's acoustic representation space. The results demonstrate that the proposed approach improves the ASR's performance by 40\% relative word error rate on LibriSpeech evaluation sets and rare-word evaluation. Furthermore, we have shown that the training strategy for the biasing modules are very crucial to be able to get better transformation of the speech representation to reflect the user’s phrases. This hypothesis is analysed and corroborated with correlation graphs between the bias embeddings among various stages of the training for different models, which effectively explore the tricks for training contextual bias for ASR.

% The success of Automatic Speech Recognition (ASR) depends on its ability to recognize time-variant rare-phrases.
Automatic Speech Recognition (ASR) still face challenges when recognizing time-variant rare-phrases.
Contextual biasing (CB) modules bias ASR model towards such contextually-relevant phrases. 
% Previous approaches performed CB by biasing the internal ASR representation and/or the output posteriors towards the relevant phrases.
During training, a list of biasing phrases are selected from a large pool of phrases following a sampling strategy. 
In this work we firstly analyse different sampling strategies to provide insights into the training of CB for ASR with correlation plots between the bias embeddings among various training stages.
Secondly, we introduce a neighbourhood attention (NA) that localizes self attention (SA) to the nearest neighbouring frames to further refine the CB output. The results show that this proposed approach provides on average a 25.84\% relative WER improvement on LibriSpeech sets and rare-word evaluation compared to the baseline.

\end{abstract}
\vspace{-2mm}

\vspace{-5mm}
\section{Introduction}
\label{sec:intro}
\vspace{-2mm}
Automatic Speech Recognition (ASR) systems have made significant advancements in recent years,  but they still face challenges when transcribing speech in diverse contexts, such as rare-words, non-native named entities etc \cite{le2021contextualized, munkhdalai2022fast, munkhdalai2023nam+}. 
The ASR performance can be improved with Contextual Biasing (CB), which incorporates contextual information into an ASR pipeline \cite{munkhdalai2022fast}.
Developing and training a reliable CB module need selecting effective samples for constructing contextual phrases lists \cite{le2021contextualized, bleeker2023approximate}. 
This demands a sampling strategy to include informative samples to effectively improve the module generalisation in real-world scenarios.
% Therefore, the sampling strategy needs to include informative examples for the module's generalization and effectiveness in real-world scenarios. 

% \footnote{\textcolor{red}{K: here you assume understanding of CB, please explain before talking about sampling}}

% In particular, the generation of appropriate contextual phrases plays a crucial role in shaping the recognition accuracy. Therefore, 

% The process of sampling contextual phrases for training the contextual biasing module poses several challenges.
Sampling contextual phrases for training the CB module poses several challenges. 
Firstly, it is crucial to include appropriate examples that can effectively bias the ASR system towards a desired context.
% \footnote{\textcolor{red}{K: what is contextual interpretations?}}
However, including useful context without being too computationally expensive at the same time is essential \cite{munkhdalai2023nam+}.
% \footnote{\textcolor{red}{K: sentence not clear}} 
Secondly, the sampling method should aim to achieve a level of generalization that prevents overfitting \cite{le2021contextualized}.
% \footnote{\textcolor{red}{K: sentence has no backing or motivation}}
% Overfitting occurs when the model becomes too specialized in the specific examples or too easy for the model to find the necessary example during the training, leading to poor performance on unseen examples. 
To mitigate this the sampling method should include negative examples that serve as distractors, i.e., phrases that do not provide any valuable contextual information.
% \footnote{\textcolor{red}{K: this point is not motivated sufficiently}} 
By incorporating distractors, the CB module learns to distinguish between relevant and irrelevant contextual cues,
% \footnote{\textcolor{red}{K: I think this is simple and easy to understand, we need similar motivation before discussing distractors}} 
improving its ability to accurately bias the ASR system \cite{zhang2022end, munkhdalai2022fast}. However, by adding positive samples and distractors, the contextual phrases list should not be excessively large. This may lead to increased computational overhead and potentially impact the overall performance of the module \cite{munkhdalai2023nam+}. Finding an optimal sampling method becomes imperative to curate a manageable yet representative set of contextual phrases.
% \footnote{\textcolor{red}{K: comes out of nowhere, rather than discussing problem and saying we need better, I suggest a formatting where we list problem of ASR -> suggest CB is a solution -> then explain CB still has x,y,z problem (distractor, large list), -> explain what solutions are available for above -> then explain your methods as solution}}  

End-to-end contextual biasing modelling approaches are gaining popularity with different model architectures, such as associative memory \cite{munkhdalai2022fast, munkhdalai2023nam+}, gated biasing \cite{alexandridis2023gated}, adapter modules \cite{jain2020contextual, wang2021light,  fox2022improving} etc. They differ from the traditional named entity correction \cite{alon2019contextual} and language model~(LM) fusion approaches \cite{ chai2019personalization, huang2020class,wang2021light} which focus on biasing at the prediction level. In this paper, we propose end-to-end biasing on the subword level with locality enhanced attention distillation over the CB output. Further, we explore sampling strategies for training the contextual biasing module and analyse performance with an embedding correlation-based framework. Our contribution is threefold:
\vspace{-3mm}
\begin{enumerate}
\setlength{\itemsep}{1pt}
  \setlength{\parskip}{0pt}
  \setlength{\parsep}{0pt}
    \item We propose a novel contextual biasing model by using local neighbourhood biases between the acoustic representation and contextual phrases, further distilling the bias attention weights. The proposed models achieve on average 25.84\% relative WER improvement with publicly available evaluation sets compare to the baseline \cite{munkhdalai2022fast,Gulati2020ConformerCT} .
    We conduct a correlation based analysis to demonstrate the superiority of the proposed models' robustness and faster convergence.    
    \item Different sampling strategies are explored to train CB for better generalisation in realistic scenarios and analysed for CB adaptation.
    \item We analyse the representation learning dynamics between different sampling strategies and models, while empirically visualising the bias correlations. To the best of our knowledge this analysis is novel in the contextual biasing research domain. 
\end{enumerate}

% The increased computational complexity associated with a larger list can also lead to a decrease in real-time processing efficiency, limiting the practicality of the contextual biasing module in resource-constrained environments.

% Thus, 
% While there have been instances where larger lists have shown improvements in ASR results, it is essential to consider the trade-off between the size of the list and the module's performance \cite{munkhdalai2023nam+}. 

% \textcolor{red}{Missing the contribution of this paper here.}

\section{Related Works}
\label{sec:backgroun}
%% this section will be mainly about contextual biasing literature review
Several papers have recognized the challenges associated with contextual phrase selection and have proposed innovative approaches to address them. Initially, the majority of research was focused on biasing/correcting the entities after the ASR model prediction as named entity correction \cite{alon2019contextual} and Finite-state transducer~(FST)/external language model fusion for Out Of Vocabulary~(OOV) words \cite{pundak2018deep, chai2019personalization, huang2020class, le2021contextualized}. 
% \cite{alon2019contextual} focuses on enhancing the recognition of proper nouns, such as names and places, within the reference transcript by leveraging phonetically similar phrases as negative examples. By doing so, the neural model is encouraged to learn more distinct representations (\textcolor{red}{NNP??}), leading to improved transcription accuracy. The core concept is the detection and emphasis of proper nouns (NNP) in the reference transcript, with the presentation of phonetically similar  phrases as negative examples.

Shallow fusion and second-pass rescoring strategies during beam search can be applied at either the word-level or the subword-level, with subword-level biasing demonstrating superior performance \cite{alon2019contextual}. Similarly word mappings \cite{huang2020class} and WordPiece representations with Weighted Finite State Transducer (WFST) at the decoder level has been proven useful \cite{le2021contextualized}.

Gradually end-to-end bias training and biasing the ASR model using adapter bias networks become more popular, which formulate it as an OOV adaptation task \cite{pundak2018deep, munkhdalai2022fast, zhang2022end, munkhdalai2023nam+, alexandridis2023gated}. The CLAS (Contextual Language Adaptation System) \cite{pundak2018deep} utilizes specific context phrases represented as word n-grams. During the inference phase, the CLAS model encounters context phrases that may be OOV terms and unseen during training. The CLAS model does not rely on specific context information during training or require meticulous tuning of rescoring weights. The NAM model \cite{munkhdalai2022fast} is a continuation of this work \cite{pundak2018deep} and uses a Neural Associative Memory (NAM) that learns and utilizes the likely word piece (WP) token following a sequence of preceding WP tokens without relying on a language model. By reconstructing previously stored patterns using partial or corrupted variances, the NAM captures the associative transition between WP sub-sequences within the same phrase. Built upon this work, \cite{munkhdalai2023nam+} introduces some improvements with a top-k search, and achieved a remarkable 16x inference speedup. Gated CB utilizes contextual adapters and dynamic activation of biasing to reduce computational cost while achieving improved data efficiency \cite{alexandridis2023gated}. In \cite{zhang2022end} a novel enhancement to the attention-based encoder-decoder (AED) model is proposed by introducing a contextual bias attention (CBA) module to leverage the context vector of the source attention in the decoder; allowing it to focus on a specific bias embedding and adding contextual bias during decoding.

% Furthermore, the posterior distributions of the connectionist temporal classification (CTC) and attention decoder are adapted based on a predefined list of bias phrases and the CBA module. 

% With the increases popularity of the ASR transducer and CTC models, CB adapters and variations of CB biasing have been incorporated into these models for hierarchical contextual biasing \cite{jain2020contextual, wang2021light,  fox2022improving, munkhdalai2023nam+, dingliwal2023personalization}. 
% The inclusion of similar pronunciation phrases and filtering out irrelevant phrases enhances the robustness of the system. Data augmentation techniques are employed for constructing the context list.

Jain et al. \cite{jain2020contextual} enhanced an RNN-T ASR system with contextual Embedding Extractor (EE), an Attention Module (AttModule), and a Biasing Module (BiasingModule). The AttModule plays a crucial role in computing attention weights for each word by utilizing the predictor output for the non-blank text history and word embeddings. The BiasingModule identifies an active subset of context words that share a common prefix with the last unfinished word in the text history, which is then utilized to bias the Joiner component leveraging the predicted text history with the current input. The CSCv3 model integrates both acoustic information and first-pass text hypotheses to perform second-pass contextual spelling correction in ASR \cite{wang2023improving}. \cite{dingliwal2023personalization} introduces a two-phase approach that incorporates early and decoder biasing techniques to enhance the performance of a conformer-CTC model in recognizing domain-specific vocabulary.

Hybrid approaches are used to bring together the best of the both worlds of the FST based biasing and end-to-end model biasing, such as spelling correction model \cite{wang2021light}, deep biasing \& shallow fusion \cite{fox2022improving},  symbolic prefix-tree search with a neural pointer generator \cite{sun2022minimising}, self-adaptive language model \cite{xu2023cb}. 

The authors in \cite{Young2020} incorporated user-specific information and derived a bias score for each word in the system's vocabulary from the training corpus to bias the prediction. A similar approach to bias user-defined words in the context of pretrained ASR models is proposed in \cite{sathyendra2022contextual}. Choosing the right contextual phrase list is crucial to train a CB module.

\vspace{-3mm}
\section{Sampling methods for bias training}
\vspace{-2mm}
\label{sec:sampling}
%\footnote{\textcolor{red}{K: is}}
The selection of an appropriate sampling method for constructing contextual phrases lists is crucial in training a robust and efficient contextual biasing module for ASR. Some of the previous work choose to bypass the sampling step and provide a static list of contextual phrases instead \cite{xu2023cb, alexandridis2023gated, pandey2023procter, fox2022improving, sun2022minimising, wang2021light, jain2020contextual}. This approach allows for a targeted selection of representative examples but may lack adaptability to unseen or evolving contexts. It may compromise the module's generalisation and performance when facing unanticipated scenarios. On the other hand, most of the papers on contextual biasing choose to have a big pool of contextual phrases and take a small set of samples for each training step \cite{munkhdalai2023nam+, bleeker2023approximate, wang2023improving}. This approach aims prevent the model from overfitting to specific samples, allowing it to adapt and generalize to new examples during the testing phase. By randomly sampling from a larger pool, the model can learn to adjust to various contextual variations and improve its robustness \cite{wang2023improving, han2022improving, zhang2022end, chang2021context, munkhdalai2022fast, huang2020class, chai2019personalization, alon2019contextual, pundak2018deep}. One significant challenge of this approach cannot guarantee that the randomly selected contextual phrases will include the necessary examples to bias the specific audio representations effectively \cite{bleeker2023approximate}. As the sampling is random, there is a possibility that certain relevant contextual cues may be missing from the training data, leading to inefficient training and potentially limiting the module's performance in real-world scenarios. Several papers have recognized the challenges associated with contextual phrase selection and have proposed innovative approaches to address this issue \cite{bleeker2023approximate, hu2023matching, chang2023dialog, munkhdalai2023nam+, sathyendra2022contextual, le2021contextualized}.

Instead of selecting a single sample from the entire pool, Chang et al. \cite{chang2021context} selectively removed target reference words from the biasing list with a 40\% probability, promoting diversity and reducing over-reliance on specific examples. They claimed to introduce variability and prevent the model from overfitting to specific examples during training \cite{chang2021context}. An alternative method \cite{zhang2022end} applies Named Entity Recognition (NER) on the reference transcripts within a given training batch. The references are then split based on whether or not they contain entity results. For the references without entities, $k$ word $n$-grams are randomly selected from each reference, where $k$ is uniformly chosen from a range from 1 to $N_{phrases}$, and $n$ is uniformly chosen from a range from 1 to $N_{order}$. The values $N_{phrases}$ and $N_{order}$ represent the maximum number of phrases that can be selected in a reference and the maximum order of a selected phrase, respectively. It achieves a more targeted and context-aware selection of bias phrases, improving the performance of the biasing module. 
% Other a unique approach to contextual phrase is presented in \cite{chang2023dialog} which performs selection by employing predefined catalogs instead of sampling from a large pool.  Rather than randomly selecting from a vast collection of phrase, The contextual phrases are categorized into distinct catalogs, and the goal is to determine the most suitable catalog for each situation. 
% In the paper authors define 3 catalogs {\it ProperName}, {\it DeviceName}, {\it DeviceLocation}. The selection process involves matching the slot string\textcolor{red}{\st{, denoted as "si,"}} in each action string with the corresponding catalog type. The slot string represents the specific attribute or category associated with the action, providing additional context for biasing. So for example, if the slot string \textcolor{red}{\st{"si"}} in the action string refers to a proper name, the method will sample phrases from the proper names catalog. By categorizing the contextual phrases into catalogs this method enables a more targeted and precise selection of phrases that are highly relevant to the given context.
Approximate nearest neighbor phrase mining~(ANN-P mining) \cite{bleeker2023approximate} aimed at selecting phrases similar to a reference phrase based on their latent space similarity in the context encoder. 

Inspired from these methods, and finding an optimal training strategy for the proposed CB model, we have explored three sampling methods for training (SMa, SMb, SMc) and one sampling method (SMd) for evaluation. 
Initially, all the training transcripts are taken and n-grams are made from those transcript. All of these n-grams are kept in a global \emph{random n-gram pool}. We have used FLAIR \cite{akbik2019flair, yamada-etal-2020-luke} model with the transcripts to extract named and rare-word entities. For each of those entities, we have collected all the possible n-grams from all the utterances, where those entities occur. The entity vs n-gram list is stored as a mapping pool \emph{entity-n-gram pool}. Let us assume our total number of context phrases are $B$. We assume the contextual phrase batch for each utterance transcript may have a different number of positive samples~(PS) and negative samples~(NS).

%\footnote{\textcolor{red}{K: what is PS}}
%\footnote{\textcolor{red}{K: not very clear what is K, PS, NS, B, etc, may be a table can help as reference}}
\textbf{SMa:} The current transcript is split into n-grams where $n$ is from one to three. From those n-grams, $k$ number of n-grams are chosen (positive samples) and shuffled randomly, and these are the corresponding context phrases for the current utterance. If $k$ is less than $B$, the remaining n-grams are sampled (negative samples) from the \emph{random n-gram pool}.

\textbf{SMb:} The name/rare-word entities are selected from the current transcript. The n-grams are made from the neighbouring words of those entities (positive samples). If the selected number of n-grams is less than $B$, the remaining n-grams are sampled from the \emph{random n-gram pool} (negative samples).

\textbf{SMc:} The name/rare-word entities are selected from the current transcript. Using those entities, $k$ n-grams are randomly selected (positive samples) and shuffled from \emph{entity-n-gram pool}. If $k$ is less than $B$, the remaining n-grams are sampled from the \emph{random n-gram pool} (negative samples).

\textbf{SMd:} The name/rare-word entities are selected from the current transcript. These $k$ words (positive samples) along with the remaining $B-k$ n-grams from  \emph{random n-gram pool} (negative samples) are selected randomly.
% \footnote{\textcolor{red}{imo, you can make a diagram showing different sampling methods so that it is visually clear}}

\vspace{-3mm}
\section{Locality Enhanced Contextual Bias Alignment for Acoustic representation}\label{sec:LECB}
\vspace{-2mm}
\begin{figure*}
  \centering
  \includegraphics[width=0.9\linewidth, height=0.36\linewidth]{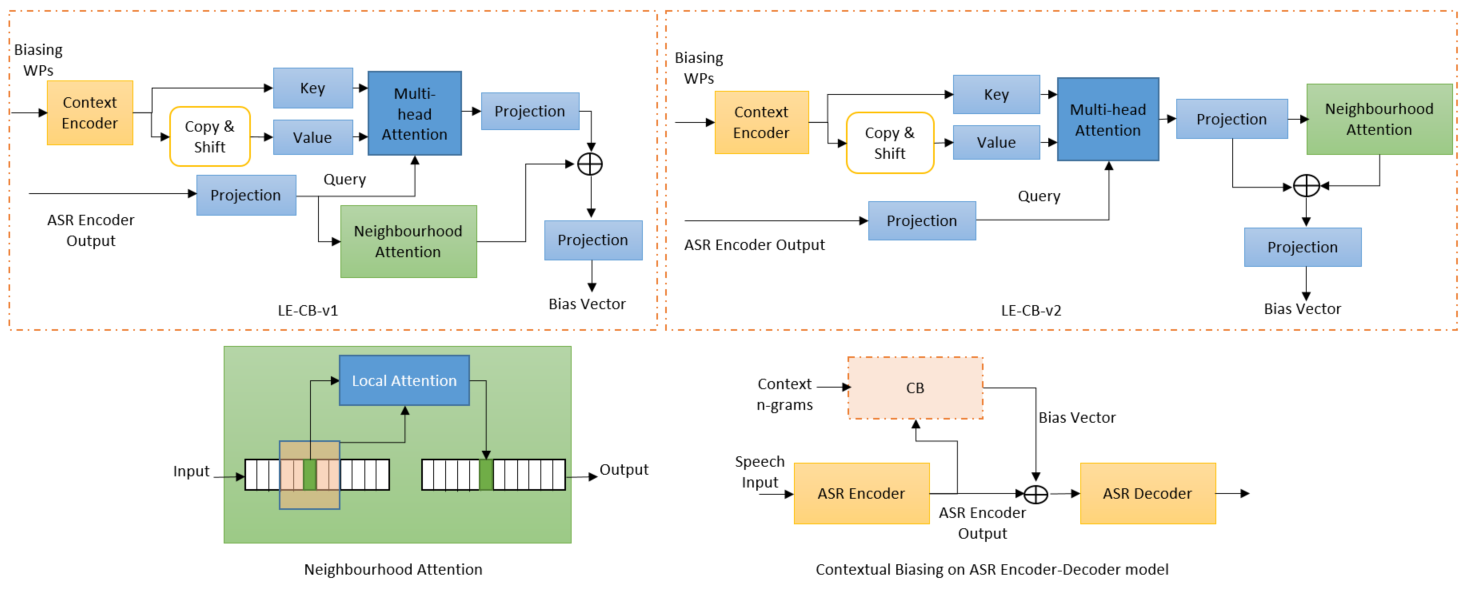}
  \caption{Locality enhanced contextual biasing}
  \label{fig:cb_architecture}
\end{figure*}

%% this section will focus on the architectures of NAM vs Amazon ( adding bias vs concatenating ) and this will explain the sparse attention for CB

% \noindent notation used for dimensions:

% \noindent context encoder embedding dim: $d$

% \noindent time dimension: $\tau$

% \noindent time dimension: $\tau$

% \noindent time features dimension: $d_{a}$

% \footnote{\textcolor{red}{K: this paragraph has same title as the paper, so does this mean your main contribution starts here? above sampling methods are not part of this paper? }}
%%maj the sampling strategies are part of the paper and they are for analysis purporse

The contextual biasing module is integrated on the baseline pre-trained ASR system to improve speech recognition accuracy by utilizing available contextual information. We use \cite{munkhdalai2022fast, chang2021context} for the base of the CB module, which models the transition between contiguous, adjacent sub-word tokens of OOV words. We further distill the context embedding using locality bounded information from the subword representation and acoustic representation using neighbourhood attention \cite{hassani2023neighborhood} with a sliding window as locality enhanced contextual biasing~(LE-CB).  
The model framework is shown in Figure \ref{fig:cb_architecture}.

\vspace{-2mm}
\subsection{Initial Bias representation learning}
\vspace{-2mm}

% In general, the module is comprised of a contextual phrase text encoder, projection linear layer to transform the audio representation, a multi-head attention mechanism that combines the audio representation feature vector with the respective contextual phrases encoded representations and a final projection layer to transform the intermediate context vector representations to the suitable output dimension to be able to merge with the context-agnostic audio features. The module accepts as input a list of $K$ contextual phrases.
Initially, the input text of each phrase that is given to the contextual biasing module is tokenized and split into subword $WP$ tokens using WordPiece algorithm \cite{wu2016googles}. The context encoder is used to extract a bi-directional contextual embedding representation for each subword unit, which results in a key matrix $K$ that stores all these vector representations. \cite{munkhdalai2022fast} proposed a Transformer-XL \cite{dai2019transformerxl} encoder without causal attention masks, but for this work we use a five layer transformer \cite{Vaswani2017AttentionIA} for learning the context subword representation. The values $V$ matrix is computed using a left-shift operation on the already computed key matrix. Each key matrix row $K_{i}$, which corresponds to the representation of token $WP_{i}$ is associated with the key representation $K_{i+1}$ from the next token $WP_{i+1}$ of the contextual phrase to store the transitions between subsequences in the contextual phrases. The resulting associated key-value pairs are passed to a multihead attention layer for retrieving
the initial bias representation. The key, value matrices are $K\in \mathbb{R}^{(N\cdot l) \times d}$, $V\in \mathbb{R}^{(N\cdot l) \times d}$, respectively, where $N$ is the total number of contextual phrases in the input, $l$ is the maximum number of tokens for each phrase and $d$ denotes the context encoder output embedding dimension. The audio feature representation vector $X \in \mathbb{R}^{\tau \times d_{a}}$, where $d_{a}$ is its feature length and $\tau$ is the time window, is used as the query vector Q for the multi head attention ~(MHA) \cite{vaswani2017attention}, before its projection to $d$ in order to match the key-value dimensionality. For the MHA operation, the traditional scaled dot product attention is utilized and calculated as follows:

\vspace{-2mm}
\begin{align}
    Att(Q,K,V) &= softmax(\frac{QK^{T}}{\sqrt{d_{k}}}V) \\
    head_{i}(Q,K,V) &= Att(QW_{i}^{Q}, KW_{i}^{K}, VW_{i}^{V})
\end{align}
Following, the MHA concatenated output $H\in \mathbb{R}^{\tau \times d}$, we use neighbourhood attention for further distilling context bias.

% a projection linear layer is used to transform the matrix dimension to $d_{a}$, before combining it with the context-agnostic initial audio feature representation $X$, as follows:
% \begin{align}
%     H_{CA}= X + H_{CB}
% \end{align}

\subsection{Neighbourhood attention over bias representation}
\label{sec:NAT}
% \footnote{\textcolor{red}{K: not clear which part is existing and which is our proposed method in this section}}
We use local biases for further attention weight distillation. Localised inductive biases are helpful for learning better representation, and MHA needs training with a huge amount of data or augmentation techniques for learning those nuanced localised biases \cite{pmlr-v139-touvron21a}. Local attention techniques such as \cite{9710580,NEURIPS2019_3416a75f, hassani2023neighborhood} improves it by using a sliding window approach for selecting regions and calculating the neighbouring regions. In our scenario, we select specific time-frames and compute attention with the neighbouring frames using neighbourhood attention~(NA) \cite{hassani2023neighborhood}, which claims to be faster and achieve transitional equivariance. From the input $H \in \mathbb{R}^{\tau \times d}$ the key-query product is computed as \cite{hassani2023neighborhood}

\begin{align*}
\textbf{A}_{i}^{k}
&= 
\begin{bmatrix}
 Q_{i}K_{\rho_{1}(i)}^{T} + B_{(i,\rho_{1}(i))}\\
 Q_{i}K_{\rho_{2}(i)}^{T} + B_{(i,\rho_{2}(i))}\\
 \cdot \\
 \cdot \\
  \cdot \\
 Q_{i}K_{\rho_{k}(i)}^{T} + B_{(i,\rho_{k}(i))}\\
\end{bmatrix},
\end{align*}
% \begin{align*}
% \textbf{V}_{i}^{k}
% &= 
% \begin{bmatrix}
%  V_{\rho_{1}(i)}\\
%  V_{\rho_{2}(i)}\\
%  \cdot \\
%  \cdot \\
%  \cdot \\
%  V_{\rho_{k}(i)}\\
% \end{bmatrix},
% \end{align*}

where $\rho_{j}(i)$ returns the $j$th nearest neighbour to $i$th input, $B_{(i,\rho_{k}(i))}$ is the relative bias for $i$th input and its $j$-th nearest neighbour, $V_{\rho_{k}(i)}$ is the projection of $H_{\rho_{k}(i)} \in \mathbb{R}^{j \times d}$ , while the total locality size is $k$. $Q_i$ is the projection of $H_i  \in \mathbb{R}^{1 \times d}$ and $K_{\rho_{k}(i)}$ is the projection of $H_{\rho_{k}(i)} \in \mathbb{R}^{j \times d}$. The neighbourhood attention $NA$ is calculated as 
\vspace{-2mm}
\begin{align}
    NA_{k}(i) = softmax \left ( \frac{\textbf{A}_{i}^{k}}{\sqrt{d}} \right )\textbf{V}_{i}^{k},
\end{align}
where the values $\textbf{V}_{i}^{k}$ consists the projection of $i$th input's $k$ nearest neighbours as $\left [ V_{\rho_{1}(i)}^{T} ... V_{\rho_{k}(i)}^{T} \right ]$ and $d$ is a scaling parameter. Two different scenarios are considered LE-CB-v1 and LE-CB-v2. In LE-CB-v1, the $H$ is $X$, which is ASR's input representation. We apply a parallel neighbourhood attention to the projection of ASR encoder's input representation and add it with the bias vector from the MHA (Equation 4). In LE-CB-v2,  the $H$ is $H_{cb}$. We apply neighbourhood attention to the biasing vector from the MHA and use a residual connection to get the final biasing vector (Equation 5). The models are explained in Figure \ref{fig:cb_architecture}. The NA representations are combined with the biasing vector with a hyperparameter $\lambda$, and projection linear layer is used to transform the matrix dimension to $d_{a}$, before combining it with the context-agnostic initial audio feature representation $X$, as follows:
\vspace{-2mm}
\begin{align}
    H_{LE-CB-v1} &= X + FFN( H_{cb}+ \lambda \cdot NA(FFN(X))) \\
    H_{LE-CB-v2} &= X + FFN( H_{cb}+ \lambda \cdot NA(FFN(H_{cb})))
\end{align} 
\vspace{-8mm}
\section{Experimental Setup}
\vspace{-4mm}
\subsection{Data}

 The LibriSpeech~\cite{7178964} corpus has been used for CB model training (in Figure \ref{fig:cb_architecture}). We have combined train-clean-100, train-clean-360, and train-other-500 into \emph{train-960} (960 hours clean speech). This combined set has been used for the training of the CB module. The \emph{dev-clean}, \emph{test-clean}, and \emph{test-other} splits have been used for validation and testing. Additionally, we have used FLAIR \cite{akbik2019flair, yamada-etal-2020-luke} to extract utterances with named entities and rare-words from \emph{dev-clean}, \emph{test-clean}, \emph{test-other} and created a \emph{libri-rare-words} evaluation set to measure rare-word performance. Furthermore, we have used our in-house \emph{apps and contacts} evaluation data with 12 users from diverse accent backgrounds with speech utterances containing names of people and applications. This evaluation set has more rare-words, and the data is also a different domain compared to LibriSpeech which consists of read-speech.
%  for measuring the model's performance on speech utterances with phone application and people's names. The motivation for using \emph{apps and contacts}\footnote{\textcolor{red}{K: this needs to be explained better}} 
\vspace{-2mm}
\subsection{Baseline}
The ASR baseline follows the transformer recipe in SpeechBrain \cite{speechbrain} for LibriSpeech\footnote{\href{https://github.com/speechbrain/speechbrain/tree/develop/recipes/LibriSpeech/ASR/transformer}{https://github.com/speechbrain/speechbrain/blob/develop/recipes/Libri-Speech/ASR/transformer}}.
The recipe implements an end-to-end transformer ASR architecture with a Conformer encoder \cite{Gulati2020ConformerCT}.
The Conformer configuration follows the Conformer (S) described in \cite{Gulati2020ConformerCT}.  The ASR baseline model is trained with LibriSpeech and CommonVoice data totalling 2k hours of data with multi-conditioned training~(MCT) \cite{rir_datasets}.

The CB baseline for this work is based on \cite{munkhdalai2022fast} which uses a pretrained ASR model along with their neural associative memory~(NAM) contextual biasing. A five layer transformer \cite{Vaswani2017AttentionIA} model is used for learning the context subword representation to bias the output of the encoder. The \emph{MHA} and the \emph{copy \& shift} are similar to \cite{munkhdalai2022fast}. Furthermore, we have evaluated without using CB to analyse the impact of contextual biasing. 

\vspace{-2mm}
 \subsection{Experimental setup}
 \vspace{-2mm}
 
 The pretrained ASR is frozen and only the CB module is trained with \emph{train-960} data. The context phrases for the training are sampled using SMa, SMb, SMc sampling strategies. All the models trained with these sampling methods are evaluated with the SMd strategy as it is the most realistic evaluation scenario for contextual biasing. SMd is suitable for on-device personalisation usecase. 
%  everything\footnote{\textcolor{red}{K: rephrase}} is  However,\footnote{\textcolor{red}{K: however followed by however}}
The $<$bias$>$ token was not used during training contrary to \cite{munkhdalai2022fast}.

% \footnote{\textcolor{red}{K: be specific rather than saying `some previous work`}} which use a bias token to prevent false positive context biasing.
\vspace{-2mm}
\subsection{Correlation Analysis}
\vspace{-2mm}
The correlation among neural embeddings for different sampling strategies and different models are calculated with SVCCA \cite{raghu2017svcca}. The motivation is to analyse the representation learning dynamics among different epochs with different models and sampling strategies. The neural embeddings are sampled from layers $l_{i}$ and $l_{j}$, and projected as $l_{i} = {z_{i}^{l_{i}},....,z_{N_{i}}^{l_{i}}}$ and $l_{j} = {z_{j}^{l_{j}},....,z_{N_{j}}^{l_{j}}}$. Singular vector decomposition is used to prune the representation dimension while reducing noise, which creates subspaces $l_{i}' \subset l_{i}$ and $l_{j}' \subset l_{j}$. Then CCA is used to find vectors $v$ and $s$ for maximising correlation $\rho$ among $l_{i}$ and $l_{j}$ as 
\vspace{-3mm}
\begin{align}
    \rho = \frac{\left< v^{T}l_{i}' , s^{T}l_{j}' \right>}{\left\|v^{T}l_{i}' \right\|\left\| s^{T}l_{j}'\right\|}
\end{align}
\vspace{-4mm}

\subsection{Evaluation}
\vspace{-2mm}
All the models used in this work have been evaluated using the SMd sampling approach from Section \ref{sec:sampling}. The context batches for evaluation with SMd contains the probable bias words and greater number of noise words / negative samples, which changes at each stage for each model. The models with \emph{LE-CB-v1} have same architecture with different weights of $\lambda$ ($\lambda = 0.5, 1.0$) in Equation 4. The results with \emph{LE-CB-v1} and \emph{LE-CB-v1} are presented in Table \ref{tab:results10}.

\vspace{-2mm}
\section{Result \& Discussion}
\SetTblrInner{colsep=2pt}
\begin{table}[]
\centering
\scriptsize
\resizebox{\linewidth}{!}{
\begin{tblr}{
  cells = {c},
  cell{3}{1} = {r=3}{},
  cell{6}{1} = {r=3}{},
  cell{9}{1} = {r=3}{},
  cell{12}{1} = {r=3}{},
  cell{15}{1} = {r=3}{},
  vlines,
  hline{1-3,6,9,12,15,18} = {-}{},
  hline{4-5,7-8,10-11,13-14,16-17} = {2-7}{},
}
\textbf{CB} & {\textbf{training}\\\textbf{sampling}} & {\textbf{dev-clean}\\\textbf{(WER)}} & {\textbf{test-clean}\\\textbf{(WER)}} & {\textbf{test-other}\\\textbf{(WER)}} & {\textbf{libri-}\\\textbf{rare-words}\\\textbf{(WER)}} & {\textbf{apps }\\\textbf{contacts}\\\textbf{(WER)}}\\
without CB & NA & 4.9 & 5.16 & 10.85 & 8.45 & 42.34\\
baseline\_NAM & SMa & 4.32 & 4.45 & 10.54 & 6.41 & 24.19\\
 & SMb & 4.49 & 4.73 & 10.49 & 7.12 & 27.7\\
 & SMc & 4.37 & 4.6 & 10.49 & 7.38 & 30.21\\
{LE-CB-v1\\$\lambda = 1.0$} & SMa & 3.87 & 4.03 & 9.46 & 6.16 & \textbf{23.08}\\
 & SMb & 3.31 & 3.49 & 8.71 & 6.19 & 26.45\\
 & SMc & 3.30 & 3.45 & 8.41 & 5.64 & 25.12\\
{LE-CB-v2\\$\lambda = 1.0$} & SMa & 3.75 & 3.97 & 9.34 & 6.63 & 23.3\\
 & SMb & \textbf{2.97} & \textbf{3.18} & \textbf{7.43} & \textbf{4.58} & 24.01\\
 & SMc & 3.54 & 3.66 & 8.95 & 6.3 & 30.63\\
{LE-CB-v1\\$\lambda = 0.5$} & SMa & 3.86 & 3.98 & 9.41 & 6.21 & 23.08\\
 & SMb & 3.08 & 3.25 & 8.13 & 5.49 & 24.21\\
 & SMc & 3.37 & 3.46 & 8.71 & 5.85 & 29.03\\
CB-C & SMa & 8.6 & 9.12 & 15.22 & 10.01 & 25.81\\
 & SMb & 3.97 & 4.12 & 9.58 & 7.46 & 24.1\\
 & SMc & 7.87 & 7.89 & 15.06 & 9.66 & 25.97
\end{tblr}
}
\caption{Contextual Biasing results with 10 context batch for each utterance and evaluation with SMd.}
\label{tab:results10}
\end{table}

%% plots for correlation analysis

%% WER comparison of SMa, SMb, SMc training and test with SMd 

%% WER comparison between attention vs sparse-attention in NAM and AMAZON model 

%% WER comparison between the NAM paper trained with the paper approach which is SMb + SMc and our approach which is SMa based.

%% weakeness of the bias token

% The results are shown in Table \ref{tab:results10}, Table \ref{tab:improvement}, and Figure \ref{fig:correlation}.\footnote{\textcolor{red}{K: this sentence is not required}} 
% \footnote{\textcolor{red}{K: sentence starts with `without`}}
Table \ref{tab:results10} shows the ASR performance on several datasets from six different models. Among them \emph{without CB} is the pretrained model without contextual biasing; \emph{baseline\_NAM} is the baseline \emph{NAM} \cite{munkhdalai2022fast} contextual biasing architecture; \emph{LE-CB-v1} \& \emph{LE-CB-v2} are the proposed models in Section \ref{sec:LECB}, equation 4 \& 5; and \emph{CB-C} is the convolution based local representation learning for the ablation study (see Section \ref{sec:ablation}). 

The results from Table \ref{tab:results10} clearly demonstrates that the proposed locality enhanced distillation of contextual bias (\emph{LE-CB-v1}, \emph{LE-CB-v2}) achieved superior results compared to the baseline. The \emph{LE-CB-v2} model emerges as the best model with relative improvement of 33.8\% WER on \emph{dev-clean}, 32.8\% WER on \emph{test-clean}, 29.1\% WER on \emph{test-other}, 35.6\% WER on \emph{libri-rare-words} and 13.32\% WER on \emph{apps \& contacts} eval set. It can be clearly seen that the results with different model variants of the \emph{LE-CB} are consistently better than the baseline model and the pretrained standalone ASR model. 

 Furthermore, the comparison between \emph{baseline\_NAM} and \emph{LE-CB} with different sampling strategies demonstrates superior training with SMb in Table \ref{tab:improvement}. The  neighbourhood attention further distils locality-based relationships in the context embeddings, as a result the \emph{LE-CB} models archive improvement compare to the baseline as well as the pretrained ASR. The average SMb improvements are over 30.51\% WER on \emph{dev-clean}, 30.09\% WER on \emph{test-clean}, 22.87\% WER on \emph{test-other}, 23.87\% WER on \emph{libri-rare-words}, 10.14\% WER on \emph{apps \& contacts}. Overall, SMb and SMc show robustness while training and evaluating CB.
 
\begin{figure}[t]
  \centering
  \includegraphics[width=1\linewidth]{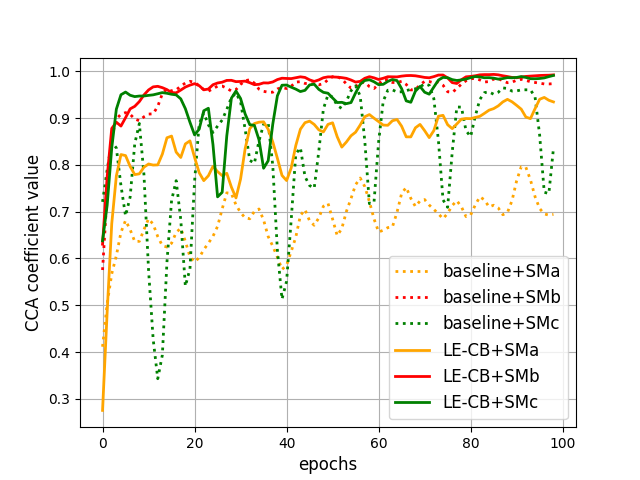}
  \vspace{-6mm}
  \caption{Contextual bias embedding representation learning analysis with epoch-wise embedding similarity}
  \label{fig:correlation}
  \vspace{-3mm}
\end{figure}

 Figure \ref{fig:correlation} shows the correlation among CB embeddings in-between successive epochs. The higher the CCA coefficient value, the closer the embeddings are. The baselines are represented with dotted curves and the \emph{LE-CB} (v1 \& v2) based models are represented with non-dotted curves. Same colors represent the same training sampling strategy. For example, red color curve in Figure \ref{fig:correlation} represents SMb. These models are trained with SMa, SMb, SMc, but epoch-wise embedding extraction in all of the models is performed with SMd ensuring a fair evaluation for all the models and dynamic context phrase list with SMd. Each of the non-dotted curves converges faster than their dotted counterpart, which clearly demonstrates that compared to the baseline, the proposed model is learning better bias representation among all the training sampling scenarios. Furthermore, the \emph{LE-CB} models are more robust because after converging, bias embedding of an utterance remains closer over the next epochs (curves are less jittery) even with different context phrase batches. It is clearly seen that training with SMb sampling method results in best representation learning and robustness. 
 
 %\footnote{\textcolor{red}{K: moving from table 1 and table 2 here is distracting}}
 %\footnote{\textcolor{red}{K: is this sentence required?}} 
 Table \ref{tab:improvement} summarises the results based on sampling strategies with the models. Among the sampling strategies, SMb and SMc produce better results than SMa for LibriSpeech eval sets (Table \ref{tab:improvement}). The models are trained and tested with different sampling strategies. Therefore, it shows that SMb and SMc produce the best generalised representation learning for contextual biasing training. However, among the evaluation sets, \emph{apps \& contacts} evaluation data is a different domain than the LibriSpeech data. SMa sampling, which is feeding the current transcript as n-grams in random order, performs reasonably well as SMc. Overall, SMc performance is weaker than SMb as Table \ref{tab:results10} and Table \ref{tab:improvement}. The possible reason can be that SMc phrases contain n-grams with the target word from all the utterances' transcripts. For example, if the current utterance is `TUMULT CRIES DOWN WITH THE BOLSHEVIKI', and the rare-word is  `BOLSHEVIKI', SMc picks the n-grams related to `BOLSHEVIKI' from the current utterance transcripts and from every other utterance where the word `BOLSHEVIKI' occurs. This causes confusion with out-of-domain evaluation data as the CB has to bias with similar neighbouring subwords while having low confidence on the mapping between acoustic-linguistic representation. One possible way to mitigate this is to use second-phase retraining with contrastive pair examples. 
 
For each positive contextual phrase, a probability between $0.7 - 1.0$ is used to determine if the phrase will remain in the batch. We have not found any significant difference in performance, however, if the probability is less than $0.7$, the performance starts to deteriorate, which explains the importance of positive context phrase samples in the context batch while training.

\begin{table}[t]
\centering
\scriptsize
\resizebox{\linewidth}{!}{
\begin{tblr}{
  cells = {c},
  cell{2}{1} = {r=3}{},
  vlines,
  hline{1-2,5} = {-}{},
  hline{3-4} = {2-7}{},
}
\textbf{CB} & {\textbf{training}\\\textbf{sampling}} & {\textbf{dev-clean}\\\textbf{(rWERR)}} & {\textbf{test-clean}\\\textbf{(rWERR)}} & {\textbf{test-other}\\\textbf{(rWERR)}} & {\textbf{libri-}\\\textbf{rare-words}\\\textbf{(rWERR)}} & {\textbf{apps \&}\\\textbf{contacts}\\\textbf{(rWERR)}}\\
LE-CB & SMa & 11.80 & 10.11 & 10.82 & -3.65 & 4.124\\
 & SMb & \textbf{30.51} & \textbf{30.09} & \textbf{22.87} & \textbf{23.87} & \textbf{10.14}\\
 & SMc & 22.12 & 23.4 & 17.16 & 19.64 & 6.45
\end{tblr}
}
\vspace{-2mm}
\caption{Average relative WER reduction (rWERR) of all the proposed models LE-CB (v1 \& v2) compared to baseline\_NAM for each sampling method with 10 context batch for each utterance.}
\label{tab:improvement}
\vspace{-5mm}
\end{table}

%\footnote{\textcolor{red}{K: sub-section can be removed and turned into a paragraph}}
\vspace{-2mm}
\subsection{Ablation Study}
\vspace{-2mm}
\label{sec:ablation}
 The neighbourhood attention applied in this paper for context bias distillation leverages the capability to calculate positional dependency on local neighbourhoods for a given time frame. From the literature, we know that convolution neural networks learn local dependencies using a moving kernel (sliding window) \cite{Gulati2020ConformerCT}. Therefore, for the ablation study, we have replaced the neighbourhood attention with a convolution layer similar to the conformer \cite{Gulati2020ConformerCT} with a skip connection. The results are shown in Table \ref{tab:results10}. The results show that the proposed models have significantly better results.
%\footnote{\textcolor{red}{K: we can provide numbers here, \% improvement}}

\vspace{-2mm}

\vspace{-2mm}
\section{Conclusion}
\vspace{-2mm}
 In this paper, we have proposed a novel contextual bias distillation technique, which uses local attention to learn neighbouring biased among subwords and acoustic representation. Furthermore, sampling strategies have been demonstrated for training contextual biasing. The best results achieved show clear improvements (on average 25.84\%) compared to the baseline with neural associative memory despite having a smaller text encoder. The exploration with the sampling strategies clearly demonstrates the representation learning and their relation to training data sampling. However, the CB module suffers loss in performance when the bias phrase list is increased as reported in other earlier research. Mitigating this performance loss will be a probable future research direction.
 \vspace{-4mm}
% \newpage

\bibliographystyle{IEEEbib}
\bibliography{refs}

\begin{thebibliography}{10}

\bibitem{le2021contextualized}
Duc Le, Mahaveer Jain, Gil Keren, Suyoun Kim, Yangyang Shi, Jay Mahadeokar,
  Julian Chan, Yuan Shangguan, Christian Fuegen, Ozlem Kalinli, et~al.,
\newblock ``Contextualized streaming end-to-end speech recognition with
  trie-based deep biasing and shallow fusion,''
\newblock {\em arXiv preprint arXiv:2104.02194}, 2021.

\bibitem{munkhdalai2022fast}
Tsendsuren Munkhdalai, Khe~Chai Sim, Angad Chandorkar, Fan Gao, Mason Chua,
  Trevor Strohman, and Fran{\c{c}}oise Beaufays,
\newblock ``Fast contextual adaptation with neural associative memory for
  on-device personalized speech recognition,''
\newblock in {\em ICASSP}, 2022, pp. 6632--6636.

\bibitem{munkhdalai2023nam+}
Tsendsuren Munkhdalai, Zelin Wu, Golan Pundak, Khe~Chai Sim, Jiayang Li, Pat
  Rondon, and Tara~N Sainath,
\newblock ``Nam+: Towards scalable end-to-end contextual biasing for adaptive
  asr,''
\newblock in {\em SLT}, 2023, pp. 190--196.

\bibitem{bleeker2023approximate}
Maurits Bleeker, Pawel Swietojanski, Stefan Braun, and Xiaodan Zhuang,
\newblock ``Approximate nearest neighbour phrase mining for contextual speech
  recognition,''
\newblock {\em arXiv preprint arXiv:2304.08862}, 2023.

\bibitem{zhang2022end}
Zhengyi Zhang and Pan Zhou,
\newblock ``End-to-end contextual asr based on posterior distribution
  adaptation for hybrid ctc/attention system,''
\newblock {\em arXiv preprint arXiv:2202.09003}, 2022.

\bibitem{alexandridis2023gated}
Anastasios Alexandridis, Kanthashree~Mysore Sathyendra, Grant Strimel,
  Feng-Ju~Claire Chang, Ariya Rastrow, Nathan Susanj, and Athanasios
  Mouchtaris,
\newblock ``Gated contextual adapters for selective contextual biasing in
  neural transducers,''
\newblock 2023.

\bibitem{jain2020contextual}
Mahaveer Jain, Gil Keren, Jay Mahadeokar, Geoffrey Zweig, Florian Metze, and
  Yatharth Saraf,
\newblock ``Contextual rnn-t for open domain asr,''
\newblock {\em arXiv preprint arXiv:2006.03411}, 2020.

\bibitem{wang2021light}
Xiaoqiang Wang, Yanqing Liu, Sheng Zhao, and Jinyu Li,
\newblock ``A light-weight contextual spelling correction model for customizing
  transducer-based speech recognition systems,''
\newblock {\em arXiv preprint arXiv:2108.07493}, 2021.

\bibitem{fox2022improving}
Jennifer~Drexler Fox and Natalie Delworth,
\newblock ``Improving contextual recognition of rare words with an alternate
  spelling prediction model,''
\newblock {\em arXiv preprint arXiv:2209.01250}, 2022.

\bibitem{alon2019contextual}
Uri Alon, Golan Pundak, and Tara~N Sainath,
\newblock ``Contextual speech recognition with difficult negative training
  examples,''
\newblock in {\em ICASSP}, 2019, pp. 6440--6444.

\bibitem{chai2019personalization}
Khe Chai~Sim, Fran{\c{c}}oise Beaufays, Arnaud Benard, Dhruv Guliani, Andreas
  Kabel, Nikhil Khare, Tamar Lucassen, Petr Zadrazil, Harry Zhang, Leif
  Johnson, et~al.,
\newblock ``Personalization of end-to-end speech recognition on mobile devices
  for named entities,''
\newblock {\em arXiv e-prints}, pp. arXiv--1912, 2019.

\bibitem{huang2020class}
Rongqing Huang, Ossama Abdel-Hamid, Xinwei Li, and Gunnar Evermann,
\newblock ``Class lm and word mapping for contextual biasing in end-to-end
  asr,''
\newblock {\em arXiv preprint arXiv:2007.05609}, 2020.

\bibitem{Gulati2020ConformerCT}
Anmol Gulati, James Qin, Chung-Cheng Chiu, Niki Parmar, Yu~Zhang, Jiahui Yu,
  Wei Han, Shibo Wang, Zhengdong Zhang, Yonghui Wu, and Ruoming Pang,
\newblock ``Conformer: Convolution-augmented transformer for speech
  recognition,''
\newblock {\em ArXiv}, vol. abs/2005.08100, 2020.

\bibitem{pundak2018deep}
Golan Pundak, Tara~N Sainath, Rohit Prabhavalkar, Anjuli Kannan, and Ding Zhao,
\newblock ``Deep context: end-to-end contextual speech recognition,''
\newblock in {\em SLT}, 2018, pp. 418--425.

\bibitem{wang2023improving}
Xiaoqiang Wang, Yanqing Liu, Jinyu Li, and Sheng Zhao,
\newblock ``Improving contextual spelling correction by external acoustics
  attention and semantic aware data augmentation,''
\newblock {\em arXiv preprint arXiv:2302.11192}, 2023.

\bibitem{dingliwal2023personalization}
Saket Dingliwal, Monica Sunkara, Srikanth Ronanki, Jeff Farris, Katrin
  Kirchhoff, and Sravan Bodapati,
\newblock ``Personalization of ctc speech recognition models,''
\newblock in {\em SLT}, 2023, pp. 302--309.

\bibitem{sun2022minimising}
Guangzhi Sun, Chao Zhang, and Philip~C Woodland,
\newblock ``Minimising biasing word errors for contextual asr with the
  tree-constrained pointer generator,''
\newblock {\em IEEE/ACM Transactions on Audio, Speech, and Language
  Processing}, vol. 31, pp. 345--354, 2022.

\bibitem{xu2023cb}
Yaoxun Xu, Baiji Liu, Qiaochu Huang, Xingchen Song, Zhiyong Wu, Shiyin Kang,
  and Helen Meng,
\newblock ``Cb-conformer: Contextual biasing conformer for biased word
  recognition,''
\newblock {\em arXiv preprint arXiv:2304.09607}, 2023.

\bibitem{Young2020}
Young~Mo Kang and Yingbo Zhou,
\newblock ``Fast and robust unsupervised contextual biasing for speech
  recognition,''
\newblock 2020.

\bibitem{sathyendra2022contextual}
Kanthashree~Mysore Sathyendra, Thejaswi Muniyappa, Feng-Ju Chang, Jing Liu,
  Jinru Su, Grant~P Strimel, Athanasios Mouchtaris, and Siegfried Kunzmann,
\newblock ``Contextual adapters for personalized speech recognition in neural
  transducers,''
\newblock in {\em ICASSP}, 2022, pp. 8537--8541.

\bibitem{pandey2023procter}
Rahul Pandey, Roger Ren, Qi~Luo, Jing Liu, Ariya Rastrow, Ankur Gandhe, Denis
  Filimonov, Grant Strimel, Andreas Stolcke, and Ivan Bulyko,
\newblock ``Procter: Pronunciation-aware contextual adapter for personalized
  speech recognition in neural transducers,''
\newblock {\em arXiv preprint arXiv:2303.17131}, 2023.

\bibitem{han2022improving}
Minglun Han, Linhao Dong, Zhenlin Liang, Meng Cai, Shiyu Zhou, Zejun Ma, and
  Bo~Xu,
\newblock ``Improving end-to-end contextual speech recognition with
  fine-grained contextual knowledge selection,''
\newblock in {\em ICASSP}, 2022, pp. 8532--8536.

\bibitem{chang2021context}
Feng-Ju Chang, Jing Liu, Martin Radfar, Athanasios Mouchtaris, Maurizio
  Omologo, Ariya Rastrow, and Siegfried Kunzmann,
\newblock ``Context-aware transformer transducer for speech recognition,''
\newblock in {\em 2021 IEEE Automatic Speech Recognition and Understanding
  Workshop (ASRU)}. IEEE, 2021, pp. 503--510.

\bibitem{hu2023matching}
Zefa Hu, Xiuyi Chen, Haoran Wu, Minglun Han, Ziyi Ni, Jing Shi, Shuang Xu, and
  Bo~Xu,
\newblock ``Matching-based term semantics pre-training for spoken patient query
  understanding,''
\newblock {\em arXiv preprint arXiv:2303.01341}, 2023.

\bibitem{chang2023dialog}
Feng-Ju Chang, Thejaswi Muniyappa, Kanthashree~Mysore Sathyendra, Kai Wei,
  Grant~P Strimel, and Ross McGowan,
\newblock ``Dialog act guided contextual adapter for personalized speech
  recognition,''
\newblock {\em arXiv preprint arXiv:2303.17799}, 2023.

\bibitem{akbik2019flair}
Alan Akbik, Tanja Bergmann, Duncan Blythe, Kashif Rasul, Stefan Schweter, and
  Roland Vollgraf,
\newblock ``{FLAIR}: An easy-to-use framework for state-of-the-art {NLP},''
\newblock in {\em {NAACL} 2019, 2019 Annual Conference of the North American
  Chapter of the Association for Computational Linguistics (Demonstrations)},
  2019, pp. 54--59.

\bibitem{yamada-etal-2020-luke}
Ikuya Yamada, Akari Asai, Hiroyuki Shindo, Hideaki Takeda, and Yuji Matsumoto,
\newblock ``{LUKE}: Deep contextualized entity representations with
  entity-aware self-attention,''
\newblock in {\em Proceedings of the 2020 Conference on Empirical Methods in
  Natural Language Processing (EMNLP)}, Online, Nov. 2020, pp. 6442--6454,
  Association for Computational Linguistics.

\bibitem{hassani2023neighborhood}
Ali Hassani, Steven Walton, Jiachen Li, Shen Li, and Humphrey Shi,
\newblock ``Neighborhood attention transformer,''
\newblock in {\em Proceedings of the IEEE/CVF Conference on Computer Vision and
  Pattern Recognition}, 2023, pp. 6185--6194.

\bibitem{wu2016googles}
Y~Wu, M~Schuster, Z~Chen, Quoc~V. Le, M~Norouzi, W~Macherey, M~Krikun, Y~Cao,
  Q~Gao, K~Macherey, J~Klingner, A~Shah, M~Johnson, Xiaobing Liu, Ł~Kaiser,
  S~Gouws, Y~Kato, T~Kudo, H~Kazawa, K~Stevens, G~Kurian, N~Patil, W~Wang,
  C~Young, J~Smith, J~Riesa, A~Rudnick, O~Vinyals, G~Corrado, M~Hughes, and
  J~Dean,
\newblock ``Google's neural machine translation system: Bridging the gap
  between human and machine translation,''
\newblock {\em ArXiv}, vol. abs/1609.08144, 2016.

\bibitem{dai2019transformerxl}
Zihang Dai, Zhilin Yang, Yiming Yang, Jaime Carbonell, Quoc~V. Le, and Ruslan
  Salakhutdinov,
\newblock ``Transformer-xl: Attentive language models beyond a fixed-length
  context,'' 2019.

\bibitem{Vaswani2017AttentionIA}
Ashish Vaswani, Noam~M. Shazeer, Niki Parmar, Jakob Uszkoreit, Llion Jones,
  Aidan~N. Gomez, Lukasz Kaiser, and Illia Polosukhin,
\newblock ``Attention is all you need,''
\newblock in {\em NIPS}, 2017.

\bibitem{vaswani2017attention}
Ashish Vaswani, Noam Shazeer, Niki Parmar, Jakob Uszkoreit, Llion Jones,
  Aidan~N Gomez, {\L}ukasz Kaiser, and Illia Polosukhin,
\newblock ``Attention is all you need,''
\newblock {\em Advances in neural information processing systems}, vol. 30,
  2017.

\bibitem{pmlr-v139-touvron21a}
Hugo Touvron, Matthieu Cord, Matthijs Douze, Francisco Massa, Alexandre
  Sablayrolles, and Herve Jegou,
\newblock ``Training data-efficient image transformers \& distillation through
  attention,''
\newblock in {\em Proceedings of the 38th International Conference on Machine
  Learning}, Marina Meila and Tong Zhang, Eds. 18--24 Jul 2021, vol. 139 of
  {\em Proceedings of Machine Learning Research}, pp. 10347--10357, PMLR.

\bibitem{9710580}
Ze~Liu, Yutong Lin, Yue Cao, Han Hu, Yixuan Wei, Zheng Zhang, Stephen Lin, and
  Baining Guo,
\newblock ``Swin transformer: Hierarchical vision transformer using shifted
  windows,''
\newblock in {\em 2021 IEEE/CVF International Conference on Computer Vision
  (ICCV)}, 2021, pp. 9992--10002.

\bibitem{NEURIPS2019_3416a75f}
Prajit Ramachandran, Niki Parmar, Ashish Vaswani, Irwan Bello, Anselm Levskaya,
  and Jon Shlens,
\newblock ``Stand-alone self-attention in vision models,''
\newblock in {\em Advances in Neural Information Processing Systems},
  H.~Wallach, H.~Larochelle, A.~Beygelzimer, F.~d\textquotesingle
  Alch\'{e}-Buc, E.~Fox, and R.~Garnett, Eds. 2019, vol.~32, Curran Associates,
  Inc.

\bibitem{7178964}
Vassil Panayotov, Guoguo Chen, Daniel Povey, and Sanjeev Khudanpur,
\newblock ``Librispeech: An asr corpus based on public domain audio books,''
\newblock in {\em ICASSP}, 2015, pp. 5206--5210.

\bibitem{speechbrain}
M~Ravanelli, T~Parcollet, P~Plantinga, A~Rouhe, S~Cornell, L~Lugosch,
  C~Subakan, N~Dawalatabad, A~Heba, J~Zhong, J~Chou, S~Yeh, S~Fu, C~Liao,
  E~Rastorgueva, F~Grondin, W~Aris, H~Na, Y~Gao, R~De Mori, and Y~Bengio,
\newblock ``{SpeechBrain}: A general-purpose speech toolkit,'' 2021,
\newblock arXiv:2106.04624.

\bibitem{rir_datasets}
T.~Ko, V.~Peddinti, D.~Povey, M.~L. Seltzer, and S.~Khudanpur,
\newblock ``A study on data augmentation of reverberant speech for robust
  speech recognition,''
\newblock in {\em ICASSP}, 2017, pp. 5220--5224.

\bibitem{raghu2017svcca}
Maithra Raghu, Justin Gilmer, Jason Yosinski, and Jascha Sohl-Dickstein,
\newblock ``Svcca: Singular vector canonical correlation analysis for deep
  learning dynamics and interpretability,''
\newblock {\em Advances in neural information processing systems}, vol. 30,
  2017.

\end{thebibliography}

\end{document}